\pdfoutput=1

\documentclass[11pt]{article}

\usepackage[final]{acl}

\usepackage{times}
\usepackage{latexsym}
\usepackage{amsmath, amssymb}
\usepackage[T1]{fontenc}

\usepackage[utf8]{inputenc}

\usepackage{microtype}

\usepackage{inconsolata}

\usepackage{graphicx}

\title{Integrating Symbolic Execution into the Fine-Tuning of Code-Generating LLMs}

\author{Marina Sakharova \and Abhinav Anand \and Mira Mezini \\
  TU Darmstadt\\}

\begin{document}
\maketitle
\begin{abstract}
Code-generating Large Language Models (LLMs) have become essential tools in modern software development, enhancing productivity and accelerating development. This paper aims to investigate the fine-tuning of code-generating LLMs using Reinforcement Learning and Direct Preference Optimization, further improving their performance. To achieve this, we enhance the training data for the reward model with the help of symbolic execution techniques, ensuring more comprehensive and objective data. With symbolic execution, we create a custom dataset that better captures the nuances in code evaluation. Our reward models, fine-tuned on this dataset, demonstrate significant improvements over the baseline, CodeRL, in estimating the quality of generated code. Our code-generating LLMs, trained with the help of reward model feedback, achieve similar results compared to the CodeRL benchmark.

\end{abstract}

\section{Introduction}
Reinforcement Learning (RL) has become one of the most powerful LLM fine-tuning techniques \cite{ouyangTrainingLanguageModels2022}. RL integrates feedback into the fine-tuning process, steering the training in the direction of human preferences. There are various approaches to applying RL to LLMs, but the general idea often consists of three steps: 
\begin{enumerate}
    \item Fine-tune a pre-trained LLM with supervised training, generate multiple answers for each given prompt and assign each answer a quality score.
    \item Use the resulting \textit{preference data} to train a reward model - an LLM that learns to produce a feedback score for a given code snippet.
    \item Generate feedback with the trained reward model and use this feedback to fine-tune the text-generating LLM.
\end{enumerate}
RL has found many applications, one of which being coding assistance \cite{leCodeRLMasteringCode2022, douStepCoderImproveCode2024, wangCompilableNeuralCode2022}. According to \citet{yu$mathcalB$CoderValueBasedDeep2024}, code generation is particularly well-suited for RL because, unlike natural language tasks, the preference data can be created automatically and more objectively through the percentage of passed unit tests.\\
However, the quality of unit test feedback is highly dependent on the test data quality \cite{bellerWhenHowWhy2015}. When human developers design test cases, they may overlook a path in the Control Flow Graph (CFG) or cover one path multiple times \cite{huangHumanErrorAnalysis2017}. These errors may result in biased feedback and, thus, incorrect RL training data.\\
Our work aims to evaluate whether \textbf{symbolic execution} improves reward-based fine-tuning of code-generating models. To achieve this, we enhance the APPS dataset \cite{hendrycksMeasuringCodingChallenge2021}, a real-world coding dataset, by augmenting it with automatically generated test cases created through symbolic execution. This technique executes code with symbolic values \cite{kingSymbolicExecutionProgram1976}, restricted to specific ranges for each control flow graph (CFG) path, ensuring that every path is covered exactly once. Symbolic execution tools analyze the CFG and generate a sample input for every path, eliminating human biases in test case creation.\\
Using the augmented APPS dataset, we fine-tune the \textit{CodeT5} model \cite{wangCodeT5IdentifierawareUnified2021} with RL, comparing its performance to \textit{CodeT5-finetuned-CodeRL} \cite{leCodeRLMasteringCode2022}, a \textit{CodeT5} version fine-tuned with RL on the original APPS that achieved SOTA performance on the MBPP benchmark \cite{austinProgramSynthesisLarge2021} at the time of its release.\\Finally, we evaluate symbolic execution for Direct Preference Optimization (DPO), a supervised alternative to RL, where the model can be trained directly on a dataset of chosen-rejected code pairs, without the usage of a reward model \cite{rafailovDirectPreferenceOptimization2024}. This addition allows us to evaluate the performance of symbolic execution under both explicit (RL) and implicit (DPO) reward settings.

\section{Related work}
There have been invented several frameworks for fine-tuning coding models with RL-based strategies. RLTF, Reinforcement Learning from Unit Test Feedback, utilizes unit test results as multi-granular feedback signals that penalize incorrect basic blocks \cite{liuRLTFReinforcementLearning2023}. PPOCoder extends unit test feedback with syntactic and semantic matching scores between generated and ground truth code \cite{shojaeeExecutionbasedCodeGeneration2023}. \citet{douStepCoderImproveCode2024} introduce StepCoder, addressing the issue of not penalizing unexecuted code by decomposing generation problems into simple sub-tasks and masking out unreached code.\\
Several recent papers introduce systems that combine symbolic execution tools and LLMs during inference. \citet{wangPythonSymbolicExecution2024} propose an LLM agent that generates execution path constraints for Python code by iteratively calling a satisfiability solver. \citet{zaharudinPosterEnhancingSymbolic2024} combine LLMs with symbolic execution tools to identify code vulnerabilities, while \citet{chenLlmBasedAutomatedModeling2024} apply both to secure medical software.\\
Although research has explored RL for fine-tuning code-generating models and integrated symbolic execution with LLM inference frameworks, little attention has been paid to combining these approaches. Specifically, the use of symbolic execution for fine-tuning code-generating models remains largely unexplored. This paper aims to bridge this gap.

\section{Methodology}
Our approach consists of two main steps: preference dataset creation and LLM fine-tuning. First, we use symbolic execution to generate test cases for APPS train tasks, produce code solutions, and rank them by performance. We then sample from the ranked codes to train \textit{CodeT5-base} \cite{wangCodeT5IdentifierawareUnified2021} as a reward model, which is subsequently used to optimize the code-generating LLM, \textit{CodeT5-large-ntp-py} \cite{leCodeRLMasteringCode2022}.

\subsection{APPS analysis}
We apply symbolic execution tools on APPS \cite{hendrycksMeasuringCodingChallenge2021} - a dataset of coding problems scraped from open-source websites. APPS consists of 5000 train and 5000 test tasks of three difficulty levels, all in Python. For each task, there are several input-output pairs available for testing. We are especially interested in test cases for training data since we use them to train the reward model on code-feedback pairs. Figure \ref{fig:train-preference} presents that 2012 out of 5000 tasks in the train set contain only one test case each. This distribution results in a percentage of passed tests being either 100\% or 0\%, leading to highly coarse and unrefined feedback. Moreover, APPS test cases were manually created by humans, which opens the possibility of overseeing an execution path \cite{huangHumanErrorAnalysis2017}. In order to extend the number of test cases and ensure the coverage of all CFG paths, we generate our custom inputs.
\subsection{Test case generation}
\begin{figure}
    \centering
    \includegraphics[width=\columnwidth]{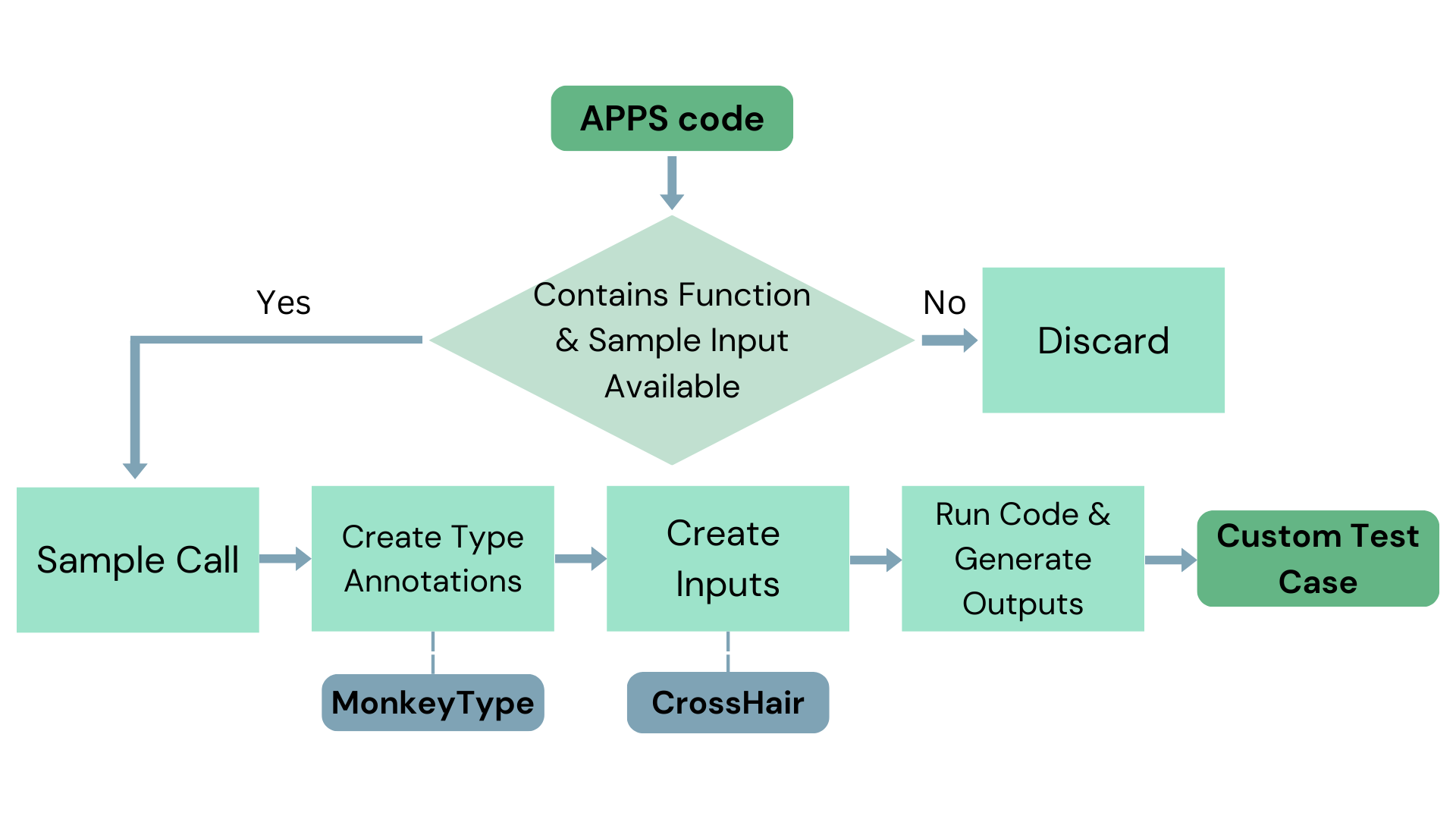}
    \caption{Test case generation pipeline.}
    \label{fig:custom-test-cases}
\end{figure}
Our input generation pipeline is presented in Figure \ref{fig:custom-test-cases}. This pipeline employs CrossHair\footnote{\url{https://github.com/pschanely/CrossHair}} - an example input generation tool for Python functions. With the help of a Satisfiability Modulo Theories solver, CrossHair explores all execution paths and finds examples and counterexamples of values.\\
To run correctly, CrossHair requires a Python function with annotated input types. Without type annotation, CrossHair outputs data of all possible types, including those irrelevant to the task. Since APPS functions lack default type hints, we use the MonkeyType annotation tool \footnote{\url{https://github.com/Instagram/MonkeyType}} to automatically infer and generate type annotations for ground truth functions based on sample input. We discard tasks that deviate from the structure of a single, standalone function and tasks that do not have any sample inputs. This filtering results in a dataset of 2402 tasks that are processed through the input generation pipeline and used for reward model training.

\subsection{Fine-tuning workflow}
\begin{figure}
    \centering
    \includegraphics[width=\columnwidth]{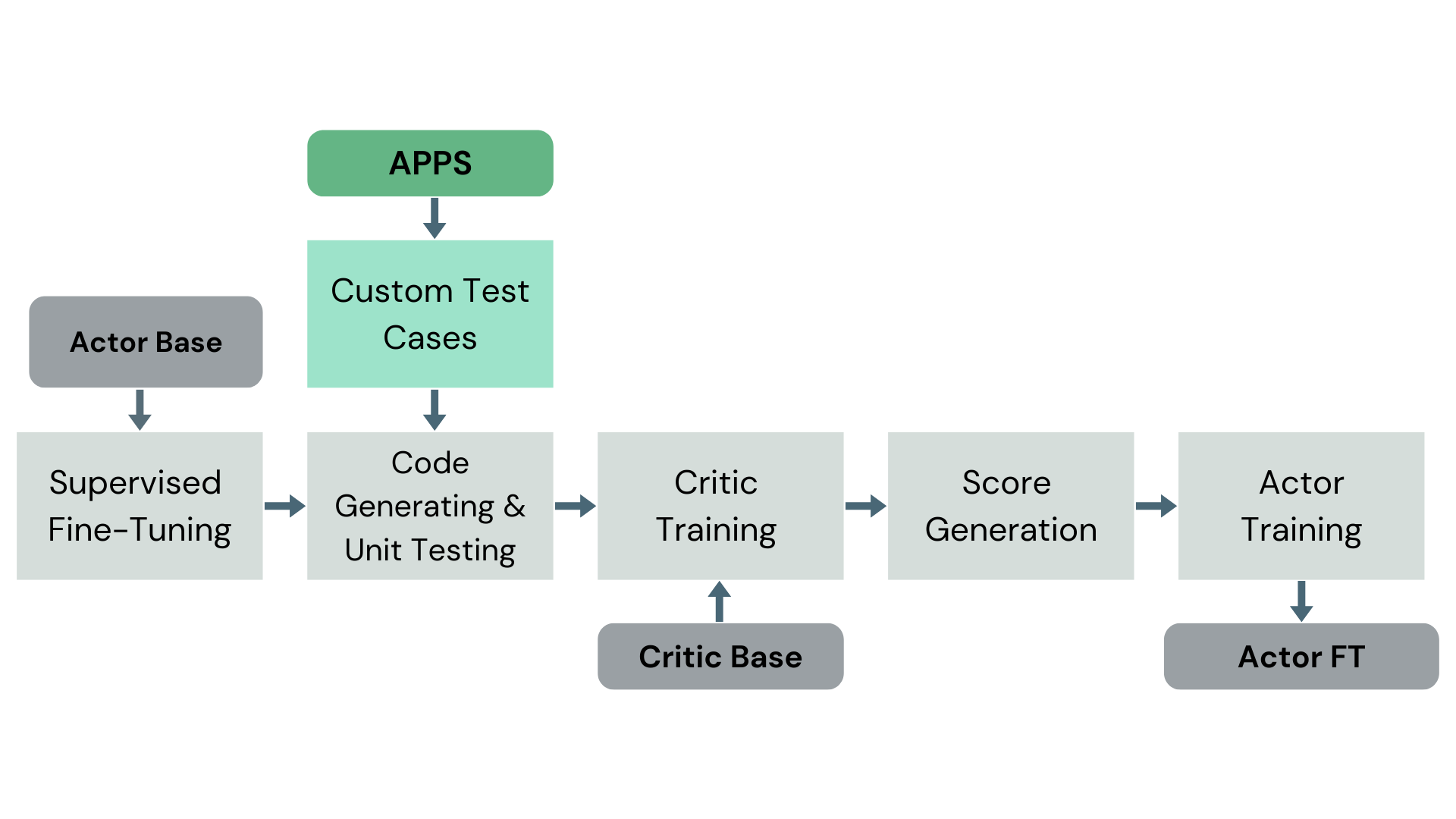}
    \caption{CodeRL training pipeline. Our pipeline extension is marked green.}
    \label{fig:coderl-workflow}
\end{figure}
Our fine-tuning pipeline relies on CodeRL \cite{leCodeRLMasteringCode2022} - a framework for RL-based LLM training. CodeRL implements an actor-critic architecture with the code-generating model as the actor and the reward model as the critic. We modify CodeRL to integrate custom test cases created with symbolic execution, as depicted in Figure \ref{fig:coderl-workflow}.\\
The training begins with a supervised warm-up phase to expose the model to NL-To-Python generation examples. We employ the original APPS training set as training data for the warm-up. A validation set, created by sampling 50\% of the original APPS test data, is used to optimize the number of warm-up epochs, with the remaining 50\% reserved for intermediate and final testing.\\
After warm-up, the LLM generates 100 codes per task for the custom training set. These codes are tested against the corresponding custom input values. For each code, the tests return a category: Compile Error, Runtime Error, (at least one) Test Failed, or Test Passed. The resulting code-feedback pairs are used to supervisely train \textit{CodeT5-base} as the critic model that classifies codes into four categories.\\
After training, the critic predicts test outcomes for each actor-generated code in the custom train set. These codes and predictions, along with ground truth solutions, are passed into the actor's training loop. Following CodeRL, we compute cross-entropy loss for ground truth data and RL loss for generated codes based on critic scores.\\
The final model is evaluated on 2,500 tasks from the APPS test set, excluding those in the validation set, and compared to the warm-up model and CodeRL baseline.
\subsection{DPO training}
In DPO, we begin the first two steps of the RL pipeline: supervised warm-up, followed by code generation for training set tasks with the new model. For each task, we select one correct solution and uniformly sample one incorrect solution to create a dataset of chosen-rejected pairs. This dataset is used to train \textit{CodeT5} with DPO trainer from the Huggingface TRL library \footnote{\url{https://huggingface.co/docs/trl/main/en/dpo_trainer}}.

\subsection{Metrics}
For evaluating actor models, we use pass@$k$ \cite{chenEvaluatingLargeLanguage2021}, the standard for measuring the performance of generated code. For each problem, if a model generates $n$ code samples and $c$ of them are correct, pass@$k$($n$, $c$, $k$) will measure the probability that at least one of the top $k$ codes passes all unit tests. The mathematical definition of this metric is presented in \ref{equation:passatk}.
\begin{equation}\label{equation:passatk}
     pass@k := \mathop{\mathbb{E}}_{\text{Problems}} \left[1 - \frac{{\binom{n-c}{k}}} {\binom{n}{k}}\right]
\end{equation}
In this paper, we use a $k$ of 5.\\
For the critic evaluation, we employ two metrics. First, we use accuracy, as the model is a classifier that predicts categorical labels. However, accuracy alone is not sufficient since it only reflects the percentage of correct predictions without considering the severity of misclassifications. The categories have an inherent order: If a code results in a compile error, it would be a less crucial mistake to predict a run-time error than code correctness. Thus, we also employ Mean Average Error, or MAE. We accordingly assign numbers from 0 to 3 to each category and calculate the absolute difference between the predicted and actual category values. This metric ensures that misclassifications involving more dissimilar categories (e.g., predicting "Test Passed" for code with a compile error) are penalized more heavily than those involving similar categories (e.g., predicting "Run-time Error" for a compile error).

\section{Experiment details}
\subsection{Critics}
We explore two training configurations to evaluate the impact of symbolic execution data:
\begin{itemize}
    \item \textbf{CodeRL-SE-critic}: Fine-tunes the existing CodeRL critic model \textit{CodeT5-finetuned-critic} \cite{leCodeRLMasteringCode2022}, enhancing it with symbolic execution inputs.
    \item \textbf{CodeT5-SE-critic}: Trains a new critic model from scratch using \textit{CodeT5-base} \cite{wangCodeT5IdentifierawareUnified2021}, the same base model used by CodeRL \cite{leCodeRLMasteringCode2022}, but with symbolic execution training data.
\end{itemize}
We train both models for one epoch using a learning rate of 2e-5. Both values are determined empirically.\\
Additionally, we evaluate the CodeRL critic model \textit{CodeT5-finetuned-critic} since the paper \cite{leCodeRLMasteringCode2022} does not provide any information about critic performance.
\subsection{Actors}
For actor training, we use \textit{CodeT5-large-ntp-py} \cite{leCodeRLMasteringCode2022}, a version of \textit{CodeT5} optimized for Python code generation tasks. We use this model because it was used as the base model for CodeRL model training. We perform two training experiments, each with one of our trained critic models, and evaluate these actors alongside the CodeRL actor. We train these models for one epoch with a learning rate of 2e-6. We determined these values empirically as well. Besides training our models, we run the inference on CodeRL actor and compare it with our results.
\subsection{DPO}
In DPO, \textit{CodeT5-large-ntp-py} is trained for one epoch, with a learning rate of 2e-6 and a $\beta$ of 0.1. $\beta$ determines how close the DPO model remains to the supervise fine-tuned model, where a smaller $\beta$ means a further deviation toward DPO loss \cite{rafailovDirectPreferenceOptimization2024}.

\section{Results}
\subsection{Enhancing APPS}
\begin{figure*}
    \centering
    \includegraphics[width=\textwidth]{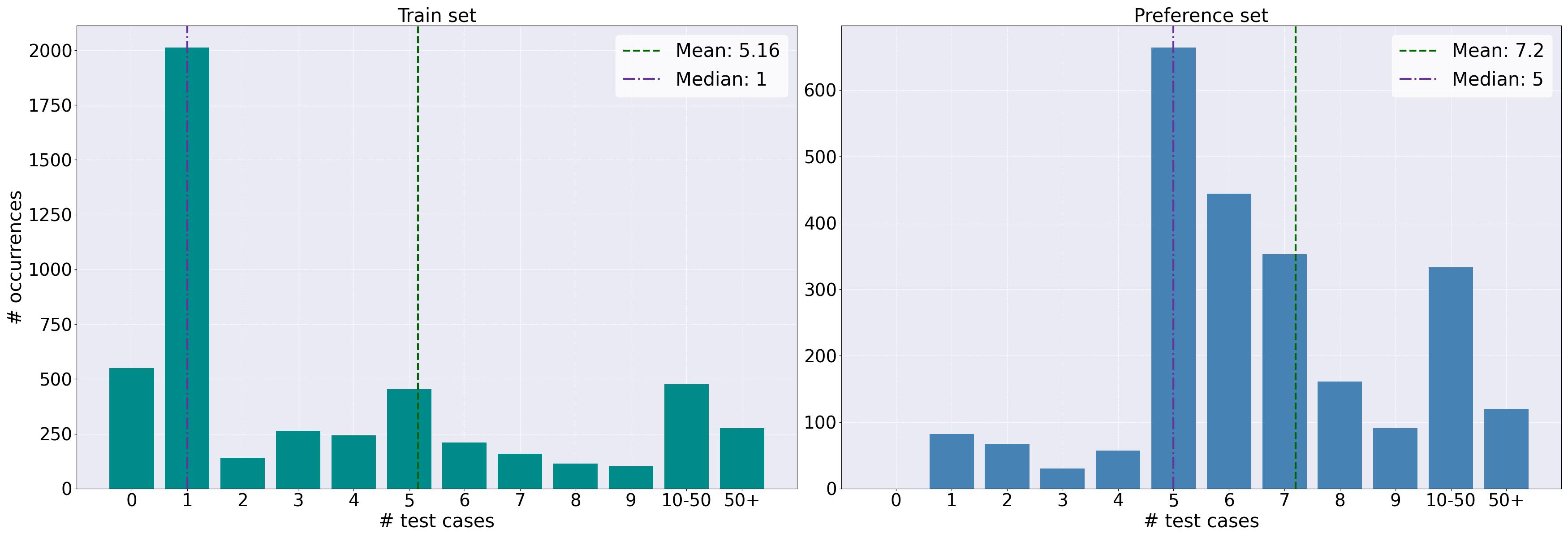}
    \caption{The distribution of test case number in the original train set (left) and the modified train set (right).}
    \label{fig:train-preference}
\end{figure*}

Figure \ref{fig:train-preference} compares the test case distributions of the original and custom symbolic execution train sets. The custom data displays a noticeable rightward skew, reflecting an increase in test case number per task.
The mean number of test cases increases from 1 to 5, and the median from 5.16 to 7.22. This observation indicates that our approach succeeded in the quantitative enhancement of the training dataset by adding more test cases.

\subsection{Critic models}
The evaluation results for the critic models are presented in Table \ref{table:critic-results}. Both of our models, \textit{CodeRL-SE-critic} and \textit{CodeT5-SE-critic}, demonstrate significant improvements over the baseline \textit{CodeT5-finetuned-critic} used in CodeRL. Among these, \textit{CodeRL-SE-critic}, a fine-tuned version of \textit{CodeT5-finetuned-critic}, achieves the highest accuracy, surpassing the original model by 37.19\%. Similarly, \textit{CodeT5-SE-critic}, which uses \textit{CodeT5-base} as its foundation, outperforms CodeRL by 11.33\%. These findings show the effectiveness of training with the symbolic execution-enhanced dataset, which positively influences the reward model's performance.
\begin{table}[!ht]
    \centering
    \begin{tabular}{llll}
    \hline
        Model &  Accuracy & MAE \\ \hline
        \textit{CodeRL-SE-critic} & \textbf{0.4250} & \textbf{0.6617} \\ 
        \textit{CodeT5-SE-critic} &  0.3449 & 0.8377 \\ 
        \textit{CodeT5-finetuned-critic} & 0.3098 & 0.9843 \\ \hline
    \end{tabular}
    \caption{Evaluation results for critic models, sorted by accuracy.}
    \label{table:critic-results}
\end{table}
\subsection{Actor models}
\begin{table*}[!ht]
    \centering
    \begin{tabular}{lllll}
    \hline
        Training method & Introductory & Interview & Competition & Total \\  \hline
        RL with \textit{CodeRL-SE-critic} & 9.42 & \textbf{3.52} & \textbf{1.91} & \textbf{4.37} \\ 
        RL (\textit{CodeT5-finetuned-CodeRL}) & \textbf{10.11} & 3.09 & 1.90 & 4.23 \\ 
        DPO  & 8.35 & 3.08 & 1.53 & 3.81\\
        RL with \textit{CodeT5-SE-critic} & 8.09 & 2.53 & 1.66 & 3.44 \\ \hline
        Supervised warm-up & 7.91&2.71&0.67&3.33 \\ 
        None (\textit{CodeT5-large-ntp-py}) & 0.00 & 0.00 & 0.00 & 0.00 \\
        \hline
    \end{tabular}
    \caption{Pass@5 results for actor models, sorted by overall performance.}
    \label{table:actor-results}
\end{table*}

The performance of \textit{CodeT5-large-ntp-py} before and after the warm-up, the actor models, and the DPO model is shown in Table \ref{table:actor-results}, divided into three difficulty levels, along with overall performance across all levels.\\
First, we can see the importance of a supervised warm-up before RL training: the results of the supervisely warmed-up model are significantly better than the base model - \textit{CodeT5-large-ntp-py}. This results in the warmed-up model being a solid base model for further fine-tuning. Moreover, we can see that all fine-tuned models, regardless of the technique and dataset used, outperform supervisely warmed-up \textit{CodeT5-large-ntp-py}. Thus, all our settings have the potential to improve LLM coding performance.\\
Nonetheless, our best actor model, \textit{RL with CodeRL-SE-critic}, achieves only a slight improvement over the CodeRL baseline \textit{CodeT5-finetuned-CodeRL}, with an overall performance gain of 0.14, measured in absolute difference. It outperforms the baseline for more complex tasks but loses for the simplest category. In contrast, our second actor, \textit{RL with CodeT5-SE-critic},  demonstrates inferior performance compared to CodeRL. Several factors could contribute to these results. In RL, if the training and evaluation distributions differ, the actor may learn to perform poorly even if the reward model scores are correct \cite{casperOpenProblemsFundamental2023}. Furthermore, RL training involves numerous hyperparameters that are challenging to optimize \cite{eimerHyperparametersReinforcementLearning2023}, and suboptimal hyperparameter tuning may have negatively impacted the model's performance.\\ 
Similarly, our DPO model also underperforms relative to CodeRL. According to \citet{xuDPOSuperiorPPO2024}, DPO models might assign disproportionately high probabilities to out-of-distribution data due to the absence of an explicit KL-divergence term. This phenomenon may explain the poor performance of DPO.\\
 While our best actor model demonstrates a slight advantage over CodeRL, the overall improvements for the actor models are notably less pronounced than those observed in the critic models. This finding challenges the intuitive expectation that a stronger reward model would lead to a more effective policy. The results raise an important question for future research: if improvements in the critic do not directly translate to better actor performance, to what extent does critic quality contribute to actor optimization compared to other factors, such as hyperparameter selection?
 
\section{Conclusion}
In this study, we investigated the intersection of fine-tuning for code-generating models and symbolic execution. By enhancing the APPS dataset with symbolic execution inputs, we ensured a solid coverage of paths within the Control Flow Graph. Using this enriched dataset, we trained two critic models that significantly outperformed the baseline - the CodeRL critic. These results indicate the high potential of using symbolic execution tools to generate training data for reward models. The enhanced coverage provided by symbolic execution enabled the reward models to access more informative and accurate training data, thereby improving their ability to evaluate a code's performance.\\
At the same time, while actor and DPO models outperformed their base models, they gained only a slight advantage over the CodeRL actor. Although our critic models predict more precise feedback, the actors stay on a similar level to CodeRL.\\
We believe that the intersection of Reinforcement Learning and symbolic execution holds significant potential for advancing code-generating models. Future work could investigate the relationship between critic performance and actor effectiveness, optimize hyperparameter configurations for actor training, and explore datasets with further programming languages or other fine-tuning tasks to achieve similar gains for actor models. With further research, we suggest that symbolic execution combined with Reinforcement Learning will enable the development of more accurate and robust coding assistants.
\section{Acknowledgements}
This research work was supported by the National Research Center for Applied Cybersecurity ATHENE.\\
I am grateful to the Zuse School ELIZA for their support throughout my academic journey. Their valuable networking opportunities have significantly contributed to my professional and personal growth.

\bibliography{acl}

\begin{thebibliography}{21}
\providecommand{\natexlab}[1]{#1}

\bibitem[{Austin et~al.(2021)Austin, Odena, Nye, Bosma, Michalewski, Dohan,
  Jiang, Cai, Terry, Le, and Sutton}]{austinProgramSynthesisLarge2021}
Jacob Austin, Augustus Odena, Maxwell Nye, Maarten Bosma, Henryk Michalewski,
  David Dohan, Ellen Jiang, Carrie Cai, Michael Terry, Quoc Le, and Charles
  Sutton. 2021.
\newblock \href {https://doi.org/10.48550/arXiv.2108.07732} {Program
  {{Synthesis}} with {{Large Language Models}}}.
\newblock \emph{Preprint}, arXiv:2108.07732.

\bibitem[{Beller et~al.(2015)Beller, Gousios, Panichella, and
  Zaidman}]{bellerWhenHowWhy2015}
Moritz Beller, Georgios Gousios, Annibale Panichella, and Andy Zaidman. 2015.
\newblock \href {https://doi.org/10.1145/2786805.2786843} {When, how, and why
  developers (do not) test in their {{IDEs}}}.
\newblock In \emph{Proceedings of the 2015 10th {{Joint Meeting}} on
  {{Foundations}} of {{Software Engineering}}}, pages 179--190, Bergamo Italy.
  ACM.

\bibitem[{Casper et~al.(2023)Casper, Davies, Shi, Gilbert, Scheurer, Rando,
  Freedman, Korbak, Lindner, Freire, Wang, Marks, Segerie, Carroll, Peng,
  Christoffersen, Damani, Slocum, Anwar, Siththaranjan, Nadeau, Michaud, Pfau,
  Krasheninnikov, Chen, Langosco, Hase, B{\i}y{\i}k, Dragan, Krueger, Sadigh,
  and {Hadfield-Menell}}]{casperOpenProblemsFundamental2023}
Stephen Casper, Xander Davies, Claudia Shi, Thomas~Krendl Gilbert,
  J{\'e}r{\'e}my Scheurer, Javier Rando, Rachel Freedman, Tomasz Korbak, David
  Lindner, Pedro Freire, Tony Wang, Samuel Marks, Charbel-Rapha{\"e}l Segerie,
  Micah Carroll, Andi Peng, Phillip Christoffersen, Mehul Damani, Stewart
  Slocum, Usman Anwar, Anand Siththaranjan, Max Nadeau, Eric~J. Michaud, Jacob
  Pfau, Dmitrii Krasheninnikov, Xin Chen, Lauro Langosco, Peter Hase, Erdem
  B{\i}y{\i}k, Anca Dragan, David Krueger, Dorsa Sadigh, and Dylan
  {Hadfield-Menell}. 2023.
\newblock \href {https://arxiv.org/abs/2307.15217} {Open {{Problems}} and
  {{Fundamental Limitations}} of {{Reinforcement Learning}} from {{Human
  Feedback}}}.
\newblock \emph{Preprint}, arXiv:2307.15217.

\bibitem[{Chen et~al.(2024)Chen, Deng, Qiu, Zhao, Lei, Song, and
  Wang}]{chenLlmBasedAutomatedModeling2024}
Jingxue Chen, Liangjun Deng, Yao Qiu, Pengbiao Zhao, Hang Lei, Jingcheng Song,
  and Xiaopei Wang. 2024.
\newblock \href {https://doi.org/10.2139/ssrn.4938953} {Llm-{{Based Automated
  Modeling}} in {{Symbolic Execution}} for {{Securing Medical Software}}}.
\newblock \emph{Preprint}, Social Science Research Network:4938953.

\bibitem[{Chen et~al.(2021)Chen, Tworek, Jun, Yuan, Pinto, Kaplan, Edwards,
  Burda, Joseph, Brockman, Ray, Puri, Krueger, Petrov, Khlaaf, Sastry, Mishkin,
  Chan, Gray, Ryder, Pavlov, Power, Kaiser, Bavarian, Winter, Tillet, Such,
  Cummings, Plappert, Chantzis, Barnes, {Herbert-Voss}, Guss, Nichol, Paino,
  Tezak, Tang, Babuschkin, Balaji, Jain, Saunders, Hesse, Carr, Leike, Achiam,
  Misra, Morikawa, Radford, Knight, Brundage, Murati, Mayer, Welinder, McGrew,
  Amodei, McCandlish, Sutskever, and Zaremba}]{chenEvaluatingLargeLanguage2021}
Mark Chen, Jerry Tworek, Heewoo Jun, Qiming Yuan, Henrique Ponde de~Oliveira
  Pinto, Jared Kaplan, Harri Edwards, Yuri Burda, Nicholas Joseph, Greg
  Brockman, Alex Ray, Raul Puri, Gretchen Krueger, Michael Petrov, Heidy
  Khlaaf, Girish Sastry, Pamela Mishkin, Brooke Chan, Scott Gray, Nick Ryder,
  Mikhail Pavlov, Alethea Power, Lukasz Kaiser, Mohammad Bavarian, Clemens
  Winter, Philippe Tillet, Felipe~Petroski Such, Dave Cummings, Matthias
  Plappert, Fotios Chantzis, Elizabeth Barnes, Ariel {Herbert-Voss},
  William~Hebgen Guss, Alex Nichol, Alex Paino, Nikolas Tezak, Jie Tang, Igor
  Babuschkin, Suchir Balaji, Shantanu Jain, William Saunders, Christopher
  Hesse, Andrew~N. Carr, Jan Leike, Josh Achiam, Vedant Misra, Evan Morikawa,
  Alec Radford, Matthew Knight, Miles Brundage, Mira Murati, Katie Mayer, Peter
  Welinder, Bob McGrew, Dario Amodei, Sam McCandlish, Ilya Sutskever, and
  Wojciech Zaremba. 2021.
\newblock \href {https://arxiv.org/abs/2107.03374} {Evaluating {{Large Language
  Models Trained}} on {{Code}}}.
\newblock \emph{Preprint}, arXiv:2107.03374.

\bibitem[{Dou et~al.(2024)Dou, Liu, Jia, Xiong, Zhou, Shen, Shan, Huang, Wang,
  Fan, Xi, Zhou, Ji, Zheng, Zhang, Huang, and
  Gui}]{douStepCoderImproveCode2024}
Shihan Dou, Yan Liu, Haoxiang Jia, Limao Xiong, Enyu Zhou, Wei Shen, Junjie
  Shan, Caishuang Huang, Xiao Wang, Xiaoran Fan, Zhiheng Xi, Yuhao Zhou, Tao
  Ji, Rui Zheng, Qi~Zhang, Xuanjing Huang, and Tao Gui. 2024.
\newblock \href {https://arxiv.org/abs/2402.01391} {{{StepCoder}}: {{Improve
  Code Generation}} with {{Reinforcement Learning}} from {{Compiler
  Feedback}}}.
\newblock \emph{Preprint}, arXiv:2402.01391.

\bibitem[{Eimer et~al.(2023)Eimer, Lindauer, and
  Raileanu}]{eimerHyperparametersReinforcementLearning2023}
Theresa Eimer, Marius Lindauer, and Roberta Raileanu. 2023.
\newblock Hyperparameters in {{Reinforcement Learning}} and {{How To Tune
  Them}}.

\bibitem[{Hendrycks et~al.(2021)Hendrycks, Basart, Kadavath, Mazeika, Arora,
  Guo, Burns, Puranik, He, Song, and
  Steinhardt}]{hendrycksMeasuringCodingChallenge2021}
Dan Hendrycks, Steven Basart, Saurav Kadavath, Mantas Mazeika, Akul Arora,
  Ethan Guo, Collin Burns, Samir Puranik, Horace He, Dawn Song, and Jacob
  Steinhardt. 2021.
\newblock \href {https://arxiv.org/abs/2105.09938} {Measuring {{Coding
  Challenge Competence With APPS}}}.
\newblock \emph{Preprint}, arXiv:2105.09938.

\bibitem[{Huang(2017)}]{huangHumanErrorAnalysis2017}
Fuqun Huang. 2017.
\newblock \href {https://doi.org/10.5772/intechopen.68392} {Human {{Error
  Analysis}} in {{Software Engineering}}}.
\newblock In \emph{Theory and {{Application}} on {{Cognitive Factors}} and
  {{Risk Management}} - {{New Trends}} and {{Procedures}}}. IntechOpen.

\bibitem[{King(1976)}]{kingSymbolicExecutionProgram1976}
James~C. King. 1976.
\newblock \href {https://doi.org/10.1145/360248.360252} {Symbolic execution and
  program testing}.
\newblock \emph{Commun. ACM}, 19(7):385--394.

\bibitem[{Le et~al.(2022)Le, Wang, Gotmare, Savarese, and
  Hoi}]{leCodeRLMasteringCode2022}
Hung Le, Yue Wang, Akhilesh~Deepak Gotmare, Silvio Savarese, and Steven C.~H.
  Hoi. 2022.
\newblock \href {https://doi.org/10.48550/arXiv.2207.01780} {{{CodeRL}}:
  {{Mastering Code Generation}} through {{Pretrained Models}} and {{Deep
  Reinforcement Learning}}}.
\newblock \emph{Preprint}, arXiv:2207.01780.

\bibitem[{Liu et~al.(2023)Liu, Zhu, Xiao, Fu, Han, Yang, and
  Ye}]{liuRLTFReinforcementLearning2023}
Jiate Liu, Yiqin Zhu, Kaiwen Xiao, Qiang Fu, Xiao Han, Wei Yang, and Deheng Ye.
  2023.
\newblock \href {https://arxiv.org/abs/2307.04349} {{{RLTF}}: {{Reinforcement
  Learning}} from {{Unit Test Feedback}}}.
\newblock \emph{Preprint}, arXiv:2307.04349.

\bibitem[{Ouyang et~al.(2022)Ouyang, Wu, Jiang, Almeida, Wainwright, Mishkin,
  Zhang, Agarwal, Slama, Ray, Schulman, Hilton, Kelton, Miller, Simens, Askell,
  Welinder, Christiano, Leike, and Lowe}]{ouyangTrainingLanguageModels2022}
Long Ouyang, Jeff Wu, Xu~Jiang, Diogo Almeida, Carroll~L. Wainwright, Pamela
  Mishkin, Chong Zhang, Sandhini Agarwal, Katarina Slama, Alex Ray, John
  Schulman, Jacob Hilton, Fraser Kelton, Luke Miller, Maddie Simens, Amanda
  Askell, Peter Welinder, Paul Christiano, Jan Leike, and Ryan Lowe. 2022.
\newblock \href {https://arxiv.org/abs/2203.02155} {Training language models to
  follow instructions with human feedback}.
\newblock \emph{Preprint}, arXiv:2203.02155.

\bibitem[{Rafailov et~al.(2024)Rafailov, Sharma, Mitchell, Ermon, Manning, and
  Finn}]{rafailovDirectPreferenceOptimization2024}
Rafael Rafailov, Archit Sharma, Eric Mitchell, Stefano Ermon, Christopher~D.
  Manning, and Chelsea Finn. 2024.
\newblock \href {https://arxiv.org/abs/2305.18290} {Direct {{Preference
  Optimization}}: {{Your Language Model}} is {{Secretly}} a {{Reward Model}}}.
\newblock \emph{Preprint}, arXiv:2305.18290.

\bibitem[{Shojaee et~al.(2023)Shojaee, Jain, Tipirneni, and
  Reddy}]{shojaeeExecutionbasedCodeGeneration2023}
Parshin Shojaee, Aneesh Jain, Sindhu Tipirneni, and Chandan~K. Reddy. 2023.
\newblock \href {https://arxiv.org/abs/2301.13816} {Execution-based {{Code
  Generation}} using {{Deep Reinforcement Learning}}}.
\newblock \emph{Preprint}, arXiv:2301.13816.

\bibitem[{Wang et~al.(2024)Wang, Liu, Chen, Li, Jin, Huang, and
  Ma}]{wangPythonSymbolicExecution2024}
Wenhan Wang, Kaibo Liu, An~Ran Chen, Ge~Li, Zhi Jin, Gang Huang, and Lei Ma.
  2024.
\newblock \href {https://doi.org/10.48550/arXiv.2409.09271} {Python {{Symbolic
  Execution}} with {{LLM-powered Code Generation}}}.
\newblock \emph{Preprint}, arXiv:2409.09271.

\bibitem[{Wang et~al.(2022)Wang, Wang, Wan, Mi, Li, Zhou, Liu, Wu, Jiang, and
  Liu}]{wangCompilableNeuralCode2022}
Xin Wang, Yasheng Wang, Yao Wan, Fei Mi, Yitong Li, Pingyi Zhou, Jin Liu, Hao
  Wu, Xin Jiang, and Qun Liu. 2022.
\newblock \href {https://doi.org/10.48550/arXiv.2203.05132} {Compilable
  {{Neural Code Generation}} with {{Compiler Feedback}}}.
\newblock \emph{Preprint}, arXiv:2203.05132.

\bibitem[{Wang et~al.(2021)Wang, Wang, Joty, and
  Hoi}]{wangCodeT5IdentifierawareUnified2021}
Yue Wang, Weishi Wang, Shafiq Joty, and Steven C.~H. Hoi. 2021.
\newblock \href {https://arxiv.org/abs/2109.00859} {{{CodeT5}}:
  {{Identifier-aware Unified Pre-trained Encoder-Decoder Models}} for {{Code
  Understanding}} and {{Generation}}}.
\newblock \emph{Preprint}, arXiv:2109.00859.

\bibitem[{Xu et~al.(2024)Xu, Fu, Gao, Ye, Liu, Mei, Wang, Yu, and
  Wu}]{xuDPOSuperiorPPO2024}
Shusheng Xu, Wei Fu, Jiaxuan Gao, Wenjie Ye, Weilin Liu, Zhiyu Mei, Guangju
  Wang, Chao Yu, and Yi~Wu. 2024.
\newblock \href {https://arxiv.org/abs/2404.10719} {Is {{DPO Superior}} to
  {{PPO}} for {{LLM Alignment}}? {{A Comprehensive Study}}}.
\newblock \emph{Preprint}, arXiv:2404.10719.

\bibitem[{Yu et~al.(2024)Yu, Tao, Chen, Sun, and
  Yang}]{yu$mathcalB$CoderValueBasedDeep2024}
Zishun Yu, Yunzhe Tao, Liyu Chen, Tao Sun, and Hongxia Yang. 2024.
\newblock \href {https://doi.org/10.48550/arXiv.2310.03173}
  {\${\textbackslash}mathcal\{\vphantom\}{{B}}\vphantom\{\}\$-{{Coder}}:
  {{Value-Based Deep Reinforcement Learning}} for {{Program Synthesis}}}.
\newblock \emph{Preprint}, arXiv:2310.03173.

\bibitem[{Zaharudin et~al.(2024)Zaharudin, Zuhaimi, and
  Shezan}]{zaharudinPosterEnhancingSymbolic2024}
Muhammad~Nabel Zaharudin, Muhammad~Haziq Zuhaimi, and Faysal~Hossain Shezan.
  2024.
\newblock Poster: {{Enhancing Symbolic Execution}} with {{LLMs}} for
  {{Vulnerability Detection}}.

\end{thebibliography}

\end{document}